\documentclass{emulateapj}
\usepackage{psfig}

\slugcomment{In press}

\shorttitle{CY 2201-3201: AN EDGE-ON SPIRAL GRAVITATIONAL LENS}
\shortauthors{CASTANDER ET AL.}

\begin{document}

\title{CXOCY J220132.8-320144: AN EDEG-ON SPIRAL GRAVITATIONAL LENS\altaffilmark{1,2}}

\author{Francisco J. Castander\altaffilmark{3,4}, Ezequiel Treister\altaffilmark{5,6,7}, Jos\'e Maza\altaffilmark{5} and Eric Gawiser\altaffilmark{6,7,8}}

\altaffiltext{1}{Partly based on observations collected at the
European Southern Observatory, Chile, under programs 72.A-509 and 74.A-0493}
\altaffiltext{2}{This paper includes data gathered with the 6.5 meter Magellan telescope located at Las campanas Observatory, Chile}
\altaffiltext{3}{Institut d'Estudis Espacials de Catalunya, Campus UAB, 08193 Bellaterra, Barcelona, Spain}
\altaffiltext{4}{Institut de Ci\`encies de l'Espai/Consejo Superor de Investigaciones Cientificas (CSIC), Campus UAB, 08193 Bellaterra, Barcelona, Spain}
\altaffiltext{5}{Departamento de Astronom\'{\i}a, Universidad de Chile, Casilla 36-D, Santiago, Chile}
\altaffiltext{6}{Department of Astronomy, Yale University, P.O. Box 208101, New Haven, CT 06520 }
\altaffiltext{7}{Yale Center for Astronomy and Astrophysics, Yale University, P.O. Box 208121, New Haven, CT 06520}
\altaffiltext{8}{National Science Foundation Astronomy and Astrophysics Postdoctoral Fellow}

\begin{abstract}
We present the CXOCY J220132.8-320144 system, which is composed of an
edge-on spiral galaxy at $z=0.32$ lensing a $z=3.9$ background
quasar. Two images of the quasar are seen. The geometry of the system
is favorable to separate the relative mass contribution of the disk
and halo in the inner parts of the galaxy. We model the system with
one elliptical mass component with the same ellipticity as the light
distribution and manage to reproduce the quasar image positions and
fluxes. We also model the system with two mass components, disk and
halo. Again, we manage to reproduce the quasar image positions and
fluxes. However, all models predict at least a third visible image
close to the disk that is not seen in our images. We speculate that
this is most likely due to extinction by the disk. We also measure the
rotational velocity of the galaxy at 2.7 disk scale radius to
be $v_c=130\pm 20$ km s$^{-1}$ from the [OII] emission lines. When
adding the rotational velocity constraint to the models, we find that
the contribution to the rotational velocity of the disk is likely to
be equal to or larger than the contribution of the halo at this
radius. The detection of the third image and a more accurate
measurement of the rotational velocity would help to set tighter
constraints on the mass distribution of this edge-on spiral galaxy.

\end{abstract}


\section{INTRODUCTION}

The current picture of galaxies is that they are composed of baryons
(stars, gas, dust) and non baryonic dark matter. While observationally the
distribution of the baryons can be studied, it is difficult to
probe how the dark matter is distributed compared to the baryons.
Traditionally, the study of galaxy dynamics has been the strongest
proof of the existence of dark matter in galaxies (e.g.,
\citealt{RF70}; \citealt{FG79}; \citealt{Rubin85}). But given the
inherent degeneracies in the inversion of dynamical data to obtain
density profiles, it is hard to measure how the dark matter is
distributed.

Another approach is to use gravitational lens systems. For strongly
lensed quasars (QSOs), the geometry and photometric properties of the
lens system depend on the projected mass inside the lensed QSO images
and therefore can be used to constrain the mass distribution.

For the particular case of spiral galaxies, generally composed of a
bulge, a disk and a dark matter halo, the rotational velocity curves
cannot disentangle the relative contributions of the different mass
components in the inner (luminous) parts of the galaxy. On the other
hand, gravitational lenses offer the possibility of doing so, especially
in the case of edge-on spirals.

There are only five spiral gravitational lenses known so far (B0218+357:
\citealt{Patnaik93}; PKS 1830-211: \citealt{Pramesh88}; Q2237+0305:
\citealt{Huchra85}; B1600+434: \citealt{Jackson95}; PMN J2004-1349:
\citealt{Winn01}). Only one of them is seen edge-on,
B1600+434. However, none of them is ideal for studying the relative mass
contribution of the different components. 

Here, we report the discovery of an edge-on spiral lens galaxy as part
of the CYDER survey, which may be the best spiral lens system to
separate the relative mass contributions of its constituents. The
Cal\'an-Yale Deep Extragalactic Research (CYDER) survey
(\citealt{Castander03b}; \citealt{Treister05}, hereafter T05) is an
optical and near-infrared imaging and spectroscopic program carried
out in archived, moderately deep {\it Chandra} fields. CXOCY
J220132.8-320144 (hereafter CY 2201-3201) is one of the faint X-ray
sources detected by CYDER in its D1 field. Optical follow-up of this
field seemed to suggest that CY 2201-3201 was an edge-on spiral with a
very bright nucleus. However, optical spectroscopy and imaging of this
source in good seeing conditions revealed its true nature. CY 2201-3201
is a lensing system, showing two images of a distant $z=3.9$ quasar
being lensed by an edge-on spiral galaxy at $z=0.32$.

The paper is structured as follows. In \S 2 we present in
chronological order the observational data gathered to characterize
this system. In \S 3 we analyze the observational constraints obtained,
and discuss them in \S 4. Finally we draw our
conclusions in \S 5. Throughout, we assume $\Omega_o=0.3$,
$\Omega_{\Lambda}=0.7$ and $H_o=70h_{70}$ km s$^{-1}$ Mpc$^{-1}$.

\begin{figure*}[ht]
\centerline{\psfig{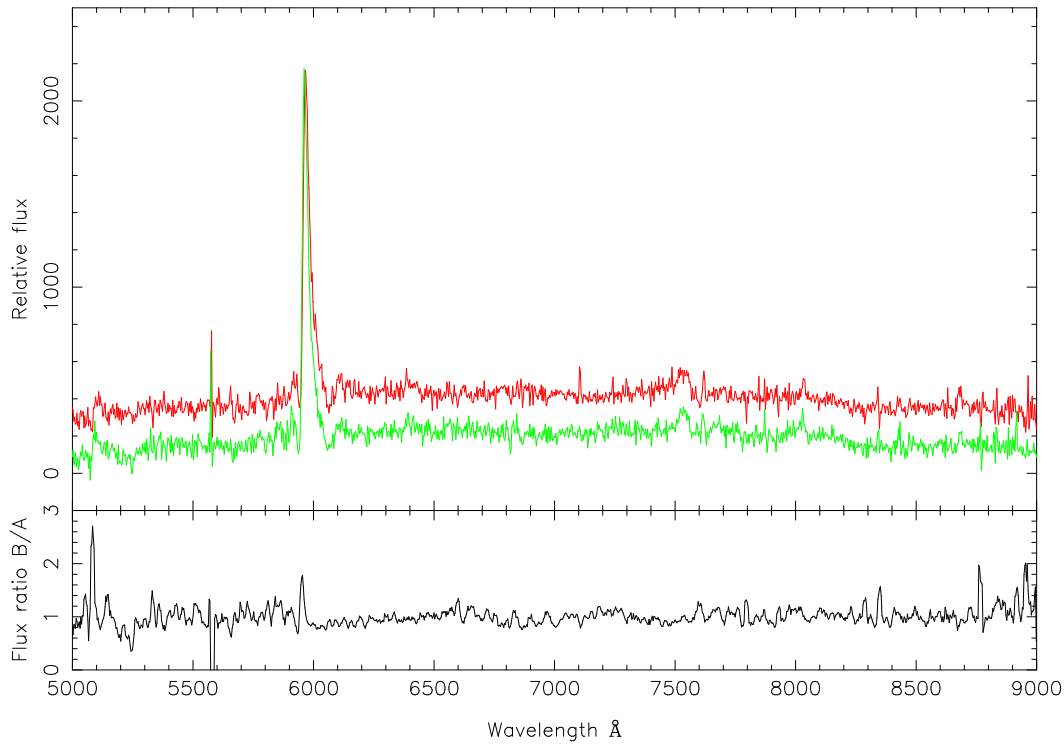}}

\caption{$Top$: VLT extracted spectra of image A ($upper line$) and B
($lower line$) of the z=3.90 lensed QSO. For displaying purposes the
spectrum of component A has been shifted vertically. $Bottom$: Flux
ratio of the two component spectra. In order to avoid contamination
from the other QSO spectrum, both QSO components extraction apertures
are narrow. Notice the small difference in apparent redshift due to a
slit width considerably wider than the seeing and the two images not
being aligned along the slit. This observational configuration makes
each image of the QSO go through a slightly different part of the
grism and therefore the same wavelength is not at the same pixel
position.\label{fig1}}

\end{figure*}

\section{OBSERVATIONS}

\subsection{X-Rays}

CY 2201-32 was observed by the $Chandra$ X-Ray Observatory with the
ACIS-I instrument for 50.16 ks in 2000 July as part of the observation
of the HCG90 field (PI Bothum). We retrieved this image from the
archive and analyzed it using standard techniques with the CIAO
package (\citealt{Castander03a}, T05).

CY 2201-32 was detected in the soft band (0.5-2.0 keV) with 11.6 $\pm$
3.9 counts and undetected in the hard band (2.0-8.0 keV). We computed
an X-ray flux of $f_x = 1.06 \times 10^{-15}$ erg cm$^{-2}$ s$^{-1}$
and X-ray luminosity of $L_x = 1.49 \times 10^{44}$ erg s$^{-1}$ in
the soft band (T05).

\subsection{CTIO Optical}

As part of the CYDER survey we imaged the CYDER D1 field (T05) with
the Cerro Tololo Inter-American Observatory (CTIO) 4m telescope using
the Mosaic II camera in 2001 August. We took images for a total of
6600 and 1800s in the $V$ and $I$ bands, reaching limiting magnitudes of
$V=26.7$ and $I=25.1$ and effective seeings of 1.0'' and 0.9'',
respectively.

CY 2201-3201 appears in these CTIO images as an edge-on spiral with a
very bright nucleus/bulge with a total measured magnitude of
$V_{Vega}=21.26$ and $I_{Vega}=19.94$.\footnote{Note that these
magnitudes differ from the ones reported in \citet{Treister05}. The
magnitudes presented here are total magnitudes while those of
\citet{Treister05} are aperture magnitudes with a relatively small
aperture missing considerable amounts of flux for this extended source.}

\subsection{VLT Spectroscopy}

On 2003 October 31, we took spectra of the X-ray optical counterparts
in the CYDER D1 field with the UT4 VLT FORS2 instrument. We used the
300V grism which gives a resolution of $R\sim 520$ (10.5 {\AA}) for
our 1'' slits. CY2201-3201 was observed in one of our masks. Given the
multi-object purpose of the masks, all slits were oriented north-south. We took
five exposures of 1800 s in seeing condition of 0.50''-0.75'' and
bright/grey sky conditions. We reduced the spectra using standard
IRAF\footnote{IRAF is distributed by the National Optical Astronomy
Observatory which is operated by AURA Inc. under contract with the
NSF.}  routines. Surprisingly, for this source we found two spectra of
a QSO at a redshift of z=3.90 in our best seeing spectra (see
Fig.~\ref{fig1}). The two QSO spectra were blended in our worse
seeing exposures. Our 30 s mask acquisition image taken with a
seeing of 0.45'' confirmed the existence of two point sources close to
the center of the edge-on spiral galaxy, and therefore confirmed the
lensing nature of the system.

\subsection{Magellan Optical Imaging}

We observed the CY 2201-3201 system with the Magellan Clay telescope
using the MagIC instrument on 2004 September 8. MagIC has a pixel
scale of 0.0691'' pixel$^{-1}$. We took a series of 300 s exposures in three
filters (SDSS $g$, $r$ and $i$) for a total of 1800, 2400, and 1800 s
in the $g$, $r$ and $i$ filters respectively.

We reduced the images using standard procedures in IRAF. The resulting
combined images have effective seeings of 0.68'', 0.65'' and 0.62'' in
$g$, $r$, and $i$, respectively.

\begin{figure}[th]
\centerline{\psfig{file=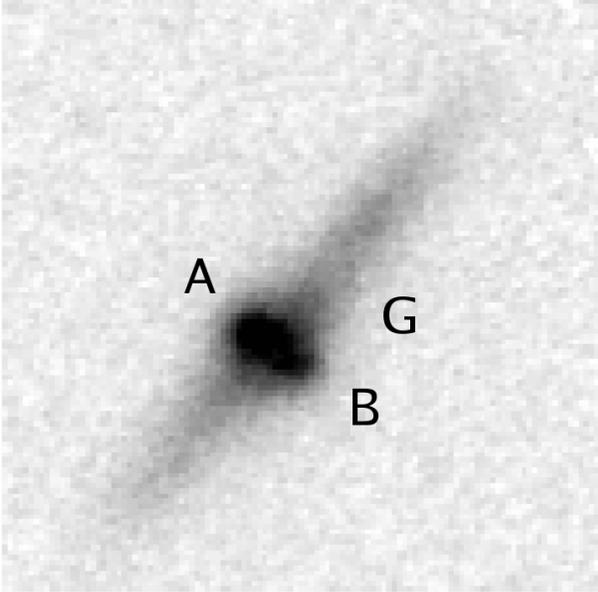,width=8cm,angle=0}}
\caption{Magellan MagIC $i$-band image of CY 2201-3201. Image size
is 8''$\times$8''. North is up and east is to the left. The QSO images are
marked A and B, and the galaxy G.\label{fig2}}
\end{figure}

Figure~\ref{fig2} shows the combined $i$-band image. The system
configuration is composed of an edge-on spiral with two images of the
QSO at each side of the disk. This is one of the expected
configurations produced by an edge-on spiral galaxy lens \citep{KK98},
except that it is missing a third image that should be located close
to the disk.

To obtain tight constraints in the lens modeling it is important to
know as accurately as possible the position and the fluxes of the
observed components. In our images the seeing is only slightly smaller
than the QSO images separation. Therefore the overlap between the QSO
images and the galaxy light distribution makes it hard to separate
their relative fluxes and determine their positions. We have thus used
a Monte Carlo Markov Chain method (MCMC; \citealt{Metropolis53};
\citealt{Hastings70}, \citealt{Gilks96}) to try to determine the
positions and fluxes (and their errors) of the three components as
accurately as possible.

We model the system with three components, two point sources (QSO\_A
and QSO\_B) and a spiral galaxy. Each point source is modeled with three
parameters: the $\alpha$ and $\delta$ of its center and its flux. The spiral
galaxy is modeled as an elliptical exponential with six parameters:
$\alpha$ and $\delta$ of the galaxy center, flux, axis ratio, position
angle of the major axis and effective radius. Our model thus has 12
parameters.

In order to obtain the best values we generate images with varying
values of these 12 parameters and convolve them with the point spread
function as given by bright nearby stars. We then compare the
resulting model image with the observed one and compute a $\chi^2$
goodness of fit. We run a MCMC to obtain the best values for 11 of the
12 parameters. We fix one parameter, the position angle of the major
axis, because the large elongation of the galaxy allows a robust
direct determination. Table~\ref{tab1} presents the parameters used to
constrain the lens model which result from a combination of the
parameters found in the three images ($g$, $r$, and $i$).  The errors
on the fluxes have been artificially increased (see below).

\begin{deluxetable}{lccc}[b]
\tablecaption{Lens Modeling Constraints of CY 2201-3201\label{tab1}}
\tablecolumns{4}
\tablewidth{0pt}
\startdata
\tableline\tableline
Parameter & QSO A & QSO B & Galaxy \\
\tableline
$\Delta \alpha$ (arcsec) & $0.00\pm0.02$ & $-0.65\pm0.03$  & $-0.48\pm0.03$  \\
$\Delta \delta$ (arcsec) & $0.00\pm0.02$ & $-0.51\pm0.03$  & $0.02\pm0.03$   \\
Relative flux            & $1.0\pm0.2$   & $0.9\pm0.2$     & ...    \\ 
Axis ratio (b/a)         & ...         & ...           & $0.11\pm0.03$   \\
Position angle (degrees)\tablenotemark{a} & ...    & ...  & $140.2$\\ 
$R$\_e (arcsec)              & ...         & ...           & $1.4\pm0.1$    \\
$V_c(2.7 r_o)$ (km s$^{-1}$) & ...         & ...          & $130\pm20$    \\
\enddata
\tablenotetext{a}{Fixed value}
\end{deluxetable}

Table~\ref{tab2} gives the measured magnitudes of the two QSO
images and the lens galaxy.

\subsection{Magellan Spectroscopy}

On 2004 September 7, 8 and 9, we took spectra of the lens galaxy using
the Magellan Clay Boller \& Chivens spectrograph (B\&C). We use a 1'' wide
slit aligned along the major axis of the lens galaxy. We use two
grisms: 600 and 1200 lines mm$^{-1}$, giving dispersions of 1.6 and
0.8 {\AA} pixel$^{-1}$, respectively. We observed the lens for 30 minutes
in 0.8'' conditions and 3 hr in 1.2'' conditions with the lower
resolution setting and for 3.5 hrs in 0.7'' conditions with the
higher resolution set-up. Unfortunately, the slit position moved with
respect to the object during the observations and the effective on-source 
exposure is shorter.

\begin{figure}[th]
\epsscale{.40}
\centerline{\psfig{file=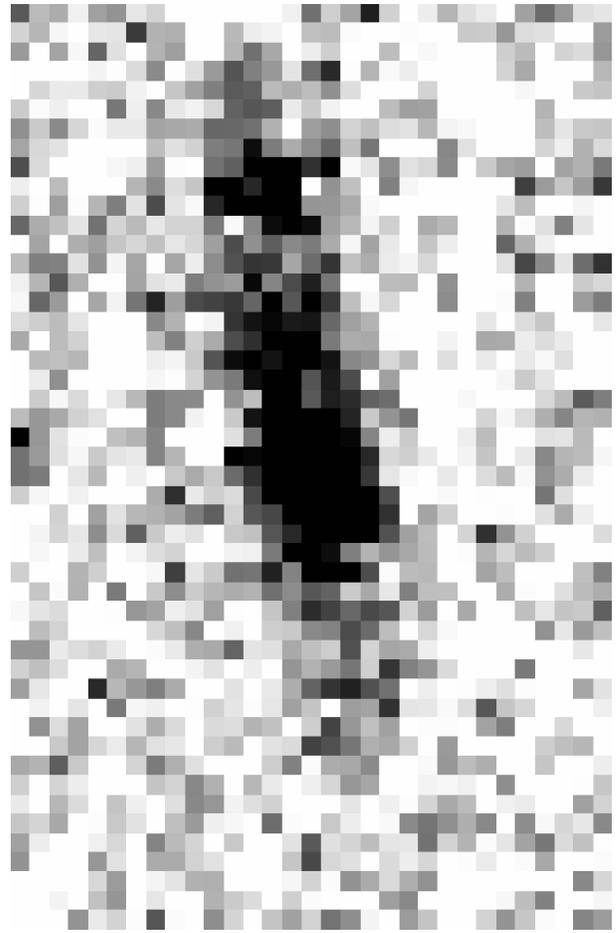,width=8cm,angle=0}}

\caption{Two-dimensional low-resolution B\&C Magellan spectrum of
CY 2201-3201. The spatial direction runs along the Y-axis and the
spectral direction along the X-axis. The sky and galaxy signals have
been subtracted.\label{fig3}}

\end{figure}

We reduced the images using standard procedures in
IRAF. 
Figure~\ref{fig3} shows the two dimensional low-resolution
spectrum at the wavelength of the observed [OII] doublet.
We measure the rotational velocity of the disk using the [OII]
doublet. We manage to detect signal out to a radius of 3.8''
equivalent to 2.7 the effective radius where the rotational velocity
appears to have already flattened. In fact for an exponential mass
distribution the maximum of the rotational velocity is at
approximately twice the effective radius \citep{BT87}. In order to
obtain a good estimate of the rotational velocity we use an MCMC.  We
model the two-dimensional spectrum assuming an exponential
distribution for the galaxy light and an arc-tangent function for the
rotational velocity curve. We generate two-dimensional spectra with
the expression

$$ I(x,y)=I_o \; exp\left(-\frac{y-y_o}{r_o}\right)\;
\frac{2}{\pi}\;v_c \; arctan\left(\frac{x-xo}{v_o}\right) $$ 
where $I_o$ is the normalization, $y_o$ and $x_o$ are the pixel
coordinates of the center of the galaxy in the two dimensional spectrum,
$r_o$ is the scale radius of the exponential profile, $v_o$ is the
``turnover'' rotational velocity radius and $v_c$ is the asymptotic
rotational velocity. We then convolve the model image with a Gaussian
of the same width of the seeing in the spatial direction and with a
Gaussian of the same width of the spectral resolution in the spectral
direction. We generate four model images: one for the low-resolution
data taken on September 7, one for the low resolution data taken on
September 9, one for the high resolution data taken on September 7 and
one for the high resolution data taken on September 8. We compare the
model images to the observed ones and compute a global $\chi^2$ of
goodness of fit. We run an MCMC with two free parameters, $I_o$ and
$v_c$, to obtain the asymptotic rotational velocity that best fit the
observations. We fix the rest of the parameters ($y_o$, $x_o$ $r_o$
and $v_o$) to the values directly measured from the two dimensional
spectra ($y_o$ and $x_o$) or the MagIC images ($r_o$). We fix $v_o$ to
the same value in pixels of $r_o$. We obtain a best value for the
asymptotic rotational velocity of $V_c = 165\pm25$ km s$^{-1}$. With
the arc-tangent rotational velocity model used, we compute a rotational
velocity at 2.7 times the scale radius of $V_c (2.7 r_o) = 130\pm20$
km s$^{-1}$. Most of the signal in the fit comes from data interior to this
radius and therefore we will use the value at this radius to compare
to the lens models that best fit the QSO image positions and fluxes.

\begin{deluxetable}{lccc}
\tablecaption{Magnitudes\tablenotemark{a} of CY 2201-3201\label{tab2}}
\tablecolumns{4}
\tablewidth{0pt}
\startdata
\tableline\tableline
Filter & QSO A & QSO B & Galaxy \\
\tableline
$g$                     & 25.06$^{+0.07}_{-0.06}$ & 25.22$^{+0.11}_{-0.10}$ & 
                        22.11$^{+0.03}_{-0.03}$ \\
$r$                     & 23.09$^{+0.05}_{-0.05}$ & 23.20$^{+0.04}_{-0.04}$ & 
                        21.06$^{+0.03}_{-0.03}$ \\
$i$xs                     & 22.62$^{+0.07}_{-0.06}$ & 22.74$^{+0.03}_{-0.03}$ & 
                        20.61$^{+0.01}_{-0.01}$ \\
$Ks$ & 18.75$^{+0.21}_{-0.18}$ & 18.96$^{+0.25}_{-0.21}$ & 
                        17.18$^{+0.29}_{-0.23}$ \\
\enddata
\tablenotetext{a}{Filters $g$, $r$, and $i$: AB system. $Ks$: Vega system}
\end{deluxetable}

In addition we measure the galaxy and QSO redshifts to be $z=0.323\pm0.001$
and $z=3.903\pm0.002$, respectively.

\subsection{NTT Near-Infrared Imaging}

We observed CY2201-3201 with the ESO New Technology Telescope (NTT) on 2004 October
30 in service mode. We took an exposure of 4185 s in the $Ks$
band. The reduced combined image provided by ESO had an effective
seeing of 0.75''. We calibrated it with standard stars.  We use the
same MCMC method used for the Magellan images to calculate the
positions and fluxes of the three modeled components in this $Ks$ image. The
calculated positions are much more uncertain than the ones calculated
with the Magellan images. The fluxes nevertheless help us constrain
the spectral energy distributions of the QSO and lens galaxy (see
Table~\ref{tab2}).

\section{LENS MODELING}

We model the CY 2201-3201 system using the positions and fluxes of the
two QSO images together with the parameters measured for the lens
galaxy.  We use the GRAVLENS
software\footnote{See http://www.physics.rutgers.edu/$\sim$ckeeton/gravlens/}
\citep{Keeton01a} developed by C. Keeton. The models used are
described in \cite{Keeton01b}. Table~\ref{tab1} summarizes the
observational constraints used for the modeling. Note that we have
artificially increased the error in the measured fluxes to allow for
micro-lensing and/or QSO variability effects. We would like to test
the hypothesis that mass is distributed similarly to light, and therefore
start modeling the system with one ellipsoidal mass component forced
to have the same ellipticity as the one observed in the light. We then
add a spherical component meant to represent the galaxy halo, which we
then allow to be slightly elongated. We also add external shear to see
the role it plays in the overall modeling. In this section we discuss
only the constraints enforced by the observed QSO positions and
fluxes. We ignore the constraints imposed by the third and sometimes
fourth images predicted but not seen and the predicted circular
velocity of the model. We discuss those constraints in the next
section.

\subsection{One Component Elliptical Mass Models}

\begin{figure*}[th]
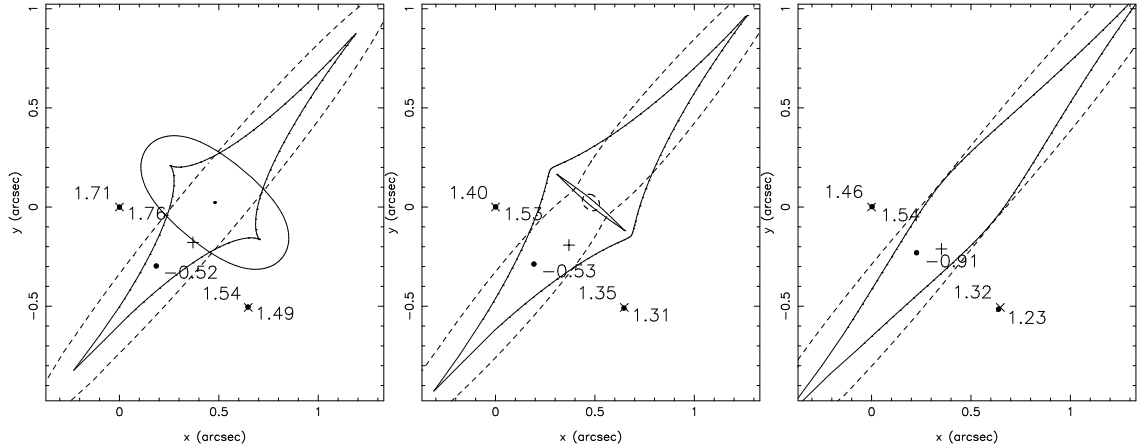

\centerline{\psfig{file=fig4a.eps,width=5cm,angle=270}\psfig{file=fig4b.eps,width=5cm,angle=270}\psfig{file=fig4c.eps,width=5cm,angle=270}}
\caption{Fits to the lens configuration (position and fluxes) for one mass component models. $Left$: SIE model. $Middle$: NFW model. $Right$: Exponential model. The solid lines
show the source plane caustics and the dashed lines the lens plane
critical curves. The plus sign ($+$) indicates the position of the QSO in the
source plane. The X symbols indicate the two observed positions of the QSO and the
dots the predicted positions. The numbers at the right and below of the
points give the predicted magnification (positive/negative values
imply positive/negative parity) and the numbers at the left and above
the Xs give the re-normalized observed magnifications.
\label{fig4}}
\end{figure*}

For simplicity, we start with a singular isothermal ellipsoid (SIE)
mass model which gives a flat rotation curve. We follow
\cite{Keeton01b} and use his ``alpha'' model. The projected mass
surface density is given by: $\kappa(\zeta)=\frac{b'}{2}\;\zeta^{-1}$,
where $\zeta=[(1-\epsilon)x^2+(1+\epsilon)y^2]^{1/2}$. $\epsilon$ is
related to the axis ratio $q$ by: $q^2=\frac{1-\epsilon}{1+\epsilon}$.
We fix the position and ellipticity of the galaxy and only allow the
normalization to vary to get the best fit to the observed QSO
positions and fluxes. Figure~\ref{fig4} shows the basic configuration
of the best fit. Three images of the QSO are produced at one side of
the rotation axis of the galaxy. This is a typical spiral galaxy lens
configuration \citep{KK98}, with the two brighter images at each side
of the disk and the fainter image closer to the
disk. Figure~\ref{fig4} shows that the position of the QSO in the
source plane (denoted by the plus sign, $+$) is very close and outside
of the radial caustic ($solid ellipsoidal line$). If the QSO were closer
to the galaxy center and inside this caustic then five images would be
produced (with the central one highly demagnified).

Next, we try an elliptical Navarro-Frenk-White (NFW) model
\citep{Navarro96,Navarro97}.  The spherical mass density in the NFW
model is given by $\rho = \frac{\rho_s}{x\;(1+x)^2}$, where
$x=r/r_s$. The projected surface mass density is $\kappa(r)=
2\;\kappa_s\;\frac{1-F(x)}{x^2-1}$, where $F(x)$ is given by equation
(48) in \cite{Keeton01b}. See also \cite{Bartelmann96}; \cite{WB00};
\cite{KS01}; \cite{GK02}; \cite{Meneghetti03}. Keeton's GRAVLENS code
defines an elliptical model from this spherical one replacing the
polar radius $r$ in the projected surface mass density expression by
the ellipse coordinate $\xi=[x^2+y^2/q^2]^{1/2}$. This model produces
the same ``disk'' configuration as the previous one. We fit the QSO
positions and fluxes allowing only the normalization and
characteristic radius of the model ($r_s$) to vary. We find that a
large range of values produce acceptable fits.  The values of the
characteristic radius and normalization that produce acceptable fits
are degenerate. The trends in this degeneracy are that, when all other
parameters are fixed, the larger the characteristic radius, the lower
the normalization and the smaller the radial caustic. That is, the
central concentration of the model dictates the size of the radial
caustic. Figure ~\ref{fig4} shows a typical configuration of this
model.

Finally, we model CY2201-3201 with an exponential mass component
\citep{Keeton01a,Keeton01b}. The projected surface mass density of
this model is given by $\kappa(\xi)=\frac{\kappa_o}{q}\;exp(-\xi/R_d)$
where $\xi$ is the same as above, $q$ is the axis ratio and $R_d$ is
the scale radius of the exponential distribution. We fix the scale
radius, the position angle and the axis ratio to the values obtained
with the light distribution (Table~\ref{tab1}) and solve for the
normalization that best reproduces the position and fluxes. The model
shows the disk configuration as before and reproduces
the position and fluxes of the two QSO images fairly well (see Fig.~\ref{fig4}).

Given that the three models produce acceptable fits, one would like to
find another constraint to differentiate between them. Comparing the
three models, we find that the position and relative magnification of
the third image depend on the concentration of the mass models. For
the most concentrated model (SIE), the position of the third image is
further away from the galaxy center and the relative magnification compared to
the brightest QSO image is the lowest (30\%). For the least
concentrated model (exponential), the position is closer to
the galaxy center and the relative magnification is the largest
(59\%).

\subsection{Two Components Mass Models (Adding a Spherical Halo)}

\begin{figure*}[th]
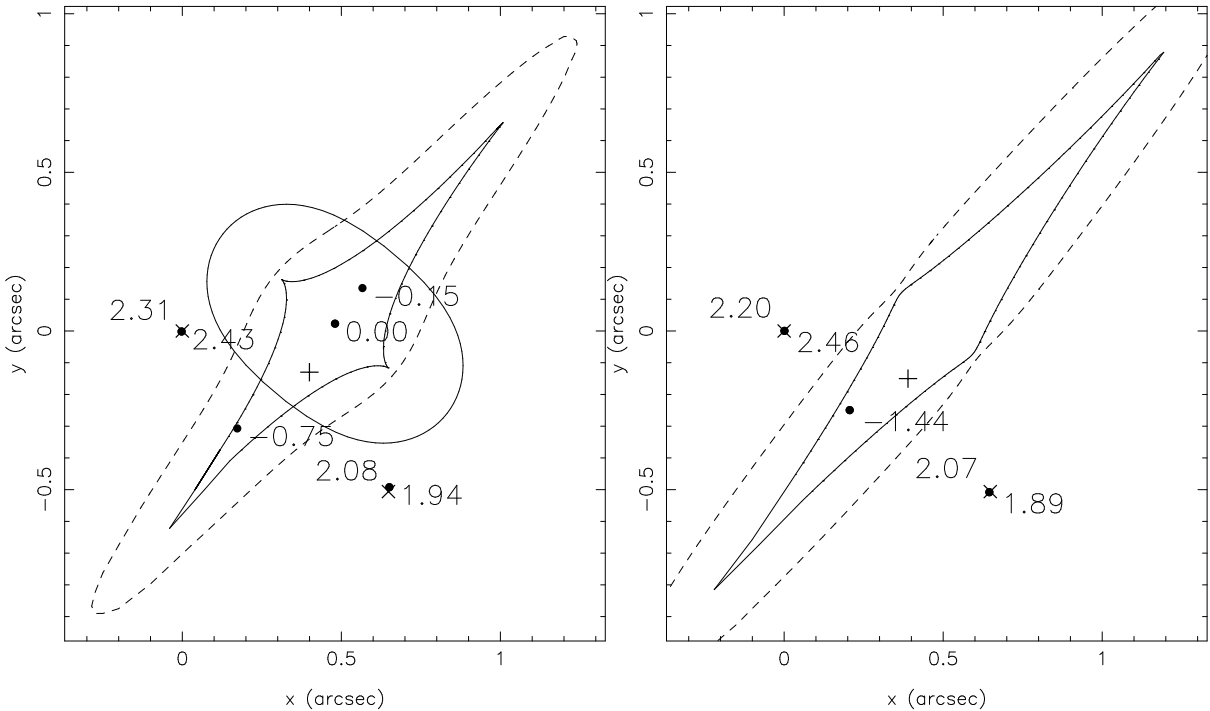

\centerline{\psfig{file=fig5a.eps,width=8cm,angle=270}\psfig{file=fig5b.eps,width=8cm,angle=270}}
\caption{Fits to the lens configuration (position and fluxes) for two-mass component models. $Left$: SIE+SIS.
$Right$: Exponential+NFW. The symbols are the same as in
Fig.~\ref{fig4}.
\label{fig5}}

\end{figure*}

We start with a simple two component model: one SIE, as in the previous
section, to which we add a single isothermal sphere (SIS). The SIS
model uses the alpha model of \cite{Keeton01b} (the same as the
SIE, see above) in which $\epsilon=0$, the axis ratio $q=1$, and
$\zeta=[x^2+y^2]^{1/2}$. We vary the normalizations of the SIS and the
SIE models, fixing all other parameters, to find the best fitting
model for the observed data. A wide range of normalizations in the
models produce acceptable fits to the position and fluxes. All
acceptable fits produce five images of the QSO (the central one
strongly de-magnified). Only models in which the rotational velocity
of the SIS is approximately 50\% larger than the rotational velocity
of the SIE do not give acceptable fits.  Figure~\ref{fig5} shows an
example of an acceptable fit in which both components have similar
rotational velocities.  Best fits are produced for models with larger
SIE contribution. In general when the contribution from the spherical
component (SIS) gets larger in the fits, then the following trends are
observed: the area enclosed by the radial caustic increases and the
area enclosed by the tangential caustic decreases; the magnification
of the QSO images gets larger and the relative magnification of the
fourth brightest image compared to the brightest gets larger. On the
other hand, the relative magnification of the third image compared to
the brightest is almost independent (staying around 30\%) of the ratio
of the SIE and SIS contributions. The fourth and fifth images
disappear at almost the point where the SIS contribution is negligible
and the model is dominated by the SIE component.

We next model the system with an expectedly more realistic mass
distribution with two components: an exponential distribution (as in
\S3.1) to account for the disk and a spherical NFW
(same as above but with the axis ratio $q=1$) representing the
halo. We search for models that best fit the data, varying the two
normalizations and the characteristic scale $r_s$ of the NFW model,
and fixing all other parameters. Again, acceptable fits to the positions
and fluxes are produced for a wide range of model parameters. The only
models that are rejected are those in which the spherical NFW
component dominates. This occurs when the rotational velocity of the
NFW is equal to or 25\% larger (depending on the concentration parameter
of the NFW) than the rotational velocity of the exponential component
at 2.7 disk scale radii. These rejected models produce five images,
with the fifth central image strongly demagnified. The only exception
is when the combination of the two mass components is such that the
QSO needs to be close to the caustic to fit the observed position and
fluxes. In this case the fifth image moves away from the center
towards the fourth image in the image plane. The trends with respect
to the relative strengths of the exponential and NFW components are
the following: The larger the NFW spherical component, the worse the
fit to the positions and the larger the magnification of the QSO
images. The relative magnification of the third brightest image
compared to the brightest image goes from $\sim$50\% when the NFW
dominates to $\sim$59\% when the exponential dominates.

If we model the halos with some ellipticity, the results present
general trends that are between the previous one mass component and
two mass components cases discussed. The most significant difference
in the case of an exponential plus elliptical NFW model is the
magnification of the resulting images. The total magnification of the
QSO images is in general lower and the relative magnification of the
third image compared to the brightest becomes larger than in the previous
cases.

\subsection{Adding the Rotational Velocity Constraint}

In the previous section we restricted ourselves to modeling the
system taking into account the positions and fluxes of the two observed
QSO images as constraints. We have also measured the rotational
velocity of the lens galaxy and can use it as an additional
constraint.

For each model we compute the rotational velocity predicted at 2.7
$r_o$, where $r_o$ is the scale radius of the exponential distribution
of the galaxy light. This is the value of the radius for which we can
measure the rotational velocity with our spectroscopic data (see
\S 2.5). All rotational velocity comparisons are performed at this
radius, neglecting the rotational velocity dependence with radius.

For the one mass component models, the best fitting SIE model to the
QSO image positions and fluxes predicts a lower rotational velocity
than observed, but consistent within the 1 $\sigma$ error. The
exponential and elliptical NFW models predict larger rotational
velocities than observed, which are consistent only at the 2 $\sigma$
level.

If we include the rotational velocity in the overall minimization of
the two component models, we find that for the SIE+SIS model the best
fit occurs when the rotational velocities of the SIS is approximately
20\% lower than the rotational velocity of the SIE. Unacceptable
models happen when the rotational velocity of the SIS is $\sim$ 5\%
larger than the rotational velocity of the SIE. For the
exponential+NFW case, we do not find acceptable models that satisfy
the three constraints: positions and fluxes of the QSO images and
rotational velocity of the galaxy. The general trend is that the
larger the relative contribution of the spherical component (halo),
the lower the rotational velocity predicted to fit the positions and
fluxes.

\section{DISCUSSION}

CY 2201-3201 is an edge-on galaxy lens. Such systems offer the
possibility of decomposing the relative mass contributions of the
disk, bulge and halo. In our optical images CY 2201-3201 appears as a
bulgeless edge-on spiral galaxy, producing two images of a background
$z=3.9$ QSO. We have therefore modeled the system with one (disk) and
two (disk+halo) mass components, neglecting the bulge.

All viable models explored produce either three or four visible
images.\footnote{In fact, there are some possible configurations where
the fifth central image, normally largely de-magnified, should also be
visible.} However, our optical images only show two images of the QSO
(Fig~\ref{fig2}). We observe the two images that are farther away
from the disk but miss the third (and in some configurations the
fourth) image located very close to the disk. We have investigated
whether dust extinction could be responsible for this missing image.
The colors of the QSO images A and B (Table~\ref{tab2}) are in fact
redder than expected for a typical QSO at that redshift.  Although low
in signal-to-noise ratioand covering a small wavelength range, the QSO
spectrum does not reveal indication of strong intrinsic absorption in
the QSO itself. One is then led to conclude that the redder colors are
due to dust extinction between the QSO and the observer. The most
probable source of attenuation comes from the disk itself. We have
computed the amount of dust extinction at the galaxy redshift
($z=0.323$) necessary to explain the observed colors assuming a mean QSO
spectrum at this redshift, no intrinsic absorption at the QSO itself
and the value of the local Galactic extinction $A_V=0.092$
\citep{Schlegel98}.
We find that the QSO image A requires an extinction by the galaxy disk
of $A_V=1.65$ and the image B an extinction of $A_V=1.55$. Images A
and B are seen at a projected distance of 0.35'' and 0.47'' (1.7 and
2.2 $h^{-1}_{70}$ kpc) of the disk plane respectively. The inferred values of
the extinction by the disk are then likely taking into account
thee projected distances to the disk. Our $i$ band image is the one
with the highest signal-to-noise detection of the QSO images. We have
inserted a third image of the QSO in our model images at the position
given by our lens models\footnote{Note that the exact position depends
on the model. In fact it can be used to distinguish between them.
Nevertheless for the purpose used here, the position can be considered
similar throughout all plausible models.} and verify through our MCMC
method the maximum flux it could have without being
significantly detected. If the ratio of expected flux to maximum
observed flux are completely due to absorption by dust at the lens
galaxy, then there are at least 3 mag of extinction more at the
position of the third image than at the brightest. Taking into account
the possible extinction at the brightest image, the third image
would suffer $A_V>4.7$ of extinction going through the lens galaxy.

In our models of the lens, we have first explored the assumption that
the mass distribution has the same ellipticity as the light. We have
tried three one-mass component models fixing the galaxy center, axis
ratio and position angle to those observed in the optical. We have fit
an SIE, an exponential and an NFW model. All three models reproduce the
two QSO images and predict a third image that is unobserved. The
predicted rotational velocities are consistent with the observed value
at the 1 $\sigma$ level for the SIE model and at the 2 $\sigma$ level
for the exponential and elliptical NFW models . We have also tried
other methods to estimate the galaxy mass. We have fit the observed
galaxy photometry with the PEGASE synthesis evolutionary models
\citep{FRV97}. We obtain an absolute magnitude of $M_r=-20.37 + 5\;
log h_{70}$,\footnote{Taking into account only the K-correction.
$M_r=-20.11 + 5\; log h_{70}$ adding an evolutionary correction.}
which is a factor 3 fainter than L$^{\star}$ \citep{Blanton03}. The
stellar mass-to-light ratio($M$/$L$) of the best fitting model is
$M/L_V\sim 4.0$, so its stellar mass would be $M \sim 3.8 \times
10^{10}$ $h_{70}^{-2}$ $M_{\odot}$. The expected rotational velocity
for such a mass at 2.7 times the scale radius is $v_c=60-85$ km
s$^{-1}$ (depending on the mass model). The stellar mass by itself is
thus insufficient to produce the observed (or predicted) rotational
velocity. If we assume the local Tully-Fisher relation \citep{TF77,
PT92, BdJ01, Verheijen01} neglecting evolution and the measured
absolute magnitude, we obtain a value for the expected rotational
velocity of $v_c\sim125-150$ km s$^{-1}$, which is consistent with the
measured rotational velocity.

We have also studied more realistic models. The galaxy does not appear
to have a bulge in the optical images, and therefore we have modelled the
system with two mass components: one for the disk and one for the
halo. We have tried an SIE+SIS model and an exponential+NFW
model. These models reproduce the positions and fluxes of the two
observed QSO images. However, the SIE+SIS model predicts four
visible images and the best fitting exponential+NFW predicts three
visible images. None of these predicted additional images are seen
in our images (see above). 

We measure a rotational velocity of $v_c=130\pm 20$ km s$^{-1}$ at 2.7
disk scale radius for the lens galaxy. The SIE+SIS model predicts
values of the rotational velocity consistent with this value for
certain combinations of the relative contributions of the SIE and SIS
components (see \S 3.3). However, the exponential+NFW model predicts
higher values of the rotational velocity if we fit the observed
position and fluxes of the QSO images. Our Magellan spectroscopic data
were obtained on three different nights with two different
settings. As explained in \S 2.5, we have four sets of same
night/grating data. The individual fits to the rotational velocity at
2.7$r_o$ for each set of data are $140\pm 40$, $160\pm 45$,$90\pm 40$,
and $130\pm 35$ (low-resolution September 7, low-resolution September
9, high-resolution September 7 and high-resolution September 8,
respectively). The relative dispersion of these values may hint at a
possible underestimation of the errors. It this were the case, the
range of allowed $v_c$ values would be larger and the exponential+NFW
model prediction would still be viable.

We have also explored other effects that can affect our modeling.  Our
system is likely to be influenced by some external shear which will
contribute to the image separation but not to the rotational velocity
of the lens galaxy. In fact, CY2201-3201 lies 7' away from
the Hickson compact group HCG 90. We have computed what would be the
external shear produced by the group assuming it is modeled with a SIE
with the same velocity dispersion as measured from the galaxy
members. HCG 90 is relatively small and the external shear induced in
the CY 2201-3201 system is negligible for our purposes. Apart from
HCG 90, CY 2201-3201 appears to be isolated and not in any group,
cluster, or obvious large-scale structure. In fact, as part of the
CYDER survey we have obtained spectra of several sources in the field
and, with the limited spectroscopic data we have, not found any sign
of a massive structure.

Another possible flaw in our modeling could be that the galaxy
center is miscalculated. 
If the galaxy center is much closer to the QSO images than the
position we have measured then most of the discussed configurations
would no longer apply and the system would display other image
configurations, which can in fact place tighter constraints on the
relative contribution of the halo and disk to the total mass budget in
the central regions of the lens galaxy.  However, the seeing and pixel
size of our images and the consistency of the galaxy center in our
different filter images make us believe that the true center if
different from the one measured should not be very far off.

\section{CONCLUSIONS}

We have presented the discovery and subsequent follow up observations
of the CY2201-3201 system composed of an edge-on spiral at $z=0.323$
splitting a background $z=3.903$ QSO into two observed images each at
opposite sides of the disk.

We have modeled the system with one (disk) and two (disk+halo) mass
components. The most likely configuration is the ``disk''configuration
with three images of the QSO one at each side of the disk and the
fainter one approximately behind the disk. There are also possible
configurations that produce four or even five (this one unlikely) observable
images. However we only observe two images. We have discussed the
possibility that the third (and fourth, if existent) image is extincted
by the disk. We estimate that an $A_V>4.7$ at th epredicted position
of the third image at the galaxy lens redshift is required to be
consistent with our observations.

We have measured the rotational velocity of the lens galaxy to be
$V_c=130\pm20$ km s$^{-1}$ at a radius 2.7 times the scale radius of
the galaxy exponential light distribution. If we use an SIE+SIS model
to fit the QSO image positions and fluxes and this value of the
rotational velocity, we find that the contribution by the SIE (disk)
to the $v_c$ at this radius is required to be the same or larger than
the contribution of the SIS (halo). If we use an exponential+NFW model
then we are unable to reproduce this value of the rotational velocity
if we fit the positions and fluxes of the QSO images. We have
speculated whether we have underestimated the error in our
determination of the $v_c$ which would make the exponential+NFW model
viable.

CY 2201-3201 is the best lensing spiral galaxy known to date that can
be used to disentangle the contributions of its different mass
components.  Unfortunately our current follow up observations are not
constraining enough to elucidate between different possible models.
More accurate image source positions, the discovery or not of the
predicted third image and the measurement of its properties and a
precise rotational velocity would make this system fulfill its
potential.

CY 2201-3201 has been awarded {\it Hubble Space Telescope} (HST) ACS time
in cycle 13. We have also been awarded more spectroscopic time at
Magellan. We expect that the higher quality images and extra spectra
will help us improve our modeling and place strong constraints on the
relative contribution of the disk and halo mass components.

\vspace*{0.2cm}

\acknowledgments

We thank C. Keeton for making public his GRAVLENS code and replying to
our questions. We thank Paul Schechter for his encouragement and
helpful discussions. We thank all observatory staff for their help
during observations. We thank the anonymous referee for his/her
comments that have helped us improve the paper. F.J.C. acknowledges
support from the Spanish Ministerio de Educaci\'on y Ciencia (MEC),
project AYA2005-09413-C02-01 with EC-FEDER funding and from the
research project 2005SGR00728 from the Generalitat de Catalunya. J.M.
gratefully acknowledges support from the Chilean Centro de
Astrof\'{\i}sica FONDAP 15010003. F.J.C. and J.M. acknowledge support from a
``convenio bilateral CSIC-Universidad de Chile''. E.G. is supported by the
National Science Foundation under grant  AST 02-01667.

\end{document}